# Coherent supercontinuum generation in tellurite glass regular lattice photonic crystal fibers

**Mariusz Klimczak,**[1,2,*] **Damian Michalik,**[1,2] **Grzegorz Stępniewski,**[1,2] **Tanvi Karpate,**[2,3] **Jarosław Cimek,**[2] **Xavier Forestier,**[2,3] **Rafał kasztelanic,**[1,2] **Dariusz Pysz,**[2] **Ryszard Stępień,**[2] **and Ryszard Buczyński**[1,2]

[1]University of Warsaw, Faculty of Physics, Pasteura 7, 02-093 Warsaw, Poland
[2]Institute of Electronic Materials Technology, Wólczyńska 133, 01-919 Warsaw, Poland
[3]Warsaw University of Technology, Faculty of Physics, Koszykowa 75, 00-662 Warsaw, Poland
mariusz.klimczak@itme.edu.pl

**Abstract:** We report on designing, fabrication and experimental characterization of highly nonlinear, tellurite glass photonic crystal fibers with engineered normal dispersion characteristics for coherent supercontinuum generation. Effectively single mode, air-hole lattice fibers, with measured, all-normal dispersion profiles as flat as -10 to -50 ps/mn/km over 1500-2400 nm wavelengths are developed and investigated. Supercontinuum spectra are measured for these fibers, with a spectral width covering 1100-2600 nm wavelengths under pumping with a robust, fixed-wavelength erbium fiber-based femtosecond laser, delivering 90 fs pulses, centered at 1560 nm with peak power below 40 kW. To the best of our knowledge, this is the first engineered microstructured fiber, which due to its high nonlinearity, enables a self-phase modulation and optical wave breaking-based supercontinuum pumped with a turn-key, 40 kW femtosecond laser at spectral widths obtainable with previous all-normal dispersion fiber designs only under pumping with systems delivering peak power in the MW range.

## 1. Introduction

Designing of specialty optical fibers for femtosecond pulse-pumped nonlinear frequency conversion allows to assume short propagation distances. Fiber length shorter than 1 cm sufficed for hyperspectral supercontinuum generation in a tellurite glass suspended core fiber, designed with the zero dispersion wavelength (ZDW) slightly blueshifted from the pump wavelength of 1550 nm [1]. Under femtosecond pumping, the pulse spectrum evolves rapidly, however such dynamics require either large initial energy or high nonlinearity. Glasses composed mainly of tellurium oxide offer the latter. This, combined with a broad transmission window from the visible to mid-infrared at around 5 μm and with high recrystallization resistance and mechanical properties superior to chalcogenide glasses, makes then an interesting alternative for the development of fibers for nonlinear optics applications. Tellurite glass nonlinear fibers have been demonstrated in context of supercontunuum generation (SG) in either step-index (SI), suspended core (SCF), tapered suspended core (T) and regular lattice photonic crystal fiber (PCF) geometries [1-24]. The general overview in Tab. 1 contains a list of fabricated tellurite fibers for SCG applications during the past 10 years updated with respect to the previous report by Désévédavy et al. [25].

Table 1. Overview of experimental results on supercontinuum generation in fibers made of various tellurite glasses.

| # | Fiber type | Glass composition *- core ** - cladding | SC pump wavelength [nm] | Peak pump power | Pump pulse duration | SC spectral coverage [nm] | Ref. and year |
|---|---|---|---|---|---|---|---|
| 1 | PCF | $75TeO_2 - 20ZnO-5Na_2O$ | 2150 | 127 kW | 120 fs | 900-2500 | [2] 2008 |
| 2 | SCF | $75TeO_2-12ZnO-5PbO-3PbF_2-5Nb_2O_5$ | 1550 | 17 kW | 110 fs | 800-4800 | [1] 2008 |
| 3 | SCF | $76.5TeO_2-6Bi_2O_3-11.5Li_2O-6ZnO$ | 1557, 1064 | 1.4 kW, 33 W | 400 fs, 15 ps | 800-2400, 800-1600 | [3] 2009 |
| 4 | PCF | $76.5TeO_2-6Bi_2O_3-6ZnO-11.5Li_2O$ | 1557 | 1.7 kW | 400 fs | 800-2400 | [4] 2009 |
| 5 | SCF | Unspecified | 1557 | 12.4 kW | 400 fs | 800-2400 | [5] 2009 |
| 6 | SCF | $76.5TeO_2–6Bi_2O_3–6ZnO–11.5Li_2O$ | 1550 | 9.4 kW | 400 fs | 470-2400 | [6] 2010 |
| 7 | SCF | $80TeO_2-10ZnO-10Na_2O$ | 1560 | ~10 kW | 100 fs | 1000-2000 | [7] 2011 |
| 8 | SI+SCF | $70TeO_2-8Li_2O-17WO_3-3MoO_3-2Nb_2O_5$* $75TeO_2-15ZnO–5Na_2O-5La_2O_3$** | 1560 | 3.8 kW | 600 fs | 1200-2400 | [8] 2011 |
| 9 | SCF | $76.5TeO_2-6Bi_2O_3 - 11.5Li_2O-6ZnO$ | 1064 | 24 W | 110 fs | 900-1600 | [9] 2011 |
| 10 | SCF | $76.5TeO_2-6Bi_2O_3-6ZnO-11.5Li_2O$ | 1064 | 387 W | 15 ps | 600-1800 | [10] 2012 |
| 11 | SCF | $80TeO_2-10ZnO-10Na_2O$ | 1745 | 7 kW | 200 fs | 750-2800 | [11] 2012 |
| 12 | SCF | Unspecified | 1560 | 3.3 W | 0.75 μs | 800-2000 | [12] 2012 |
| 13 | SCF | Unspecified | 1560 | 37 kW | 200 fs | 1400-2400 | [13] 2012 |
| 14 | PCF | $65TeO_2-28WO_3-5Na_2O-2Nb_2O_5$ | 1580 | 230 kW | 150 fs | 800-2500 | [14] 2013 |
| 15 | SCF | Unspecified | 1030 | 28-380 W | 46 fs | 750-1600 | [15] 2013 |
| 16 | SCF | $80TeO_2-(10-x)ZnO-xZnF_2–10Na_2O$, x=0.5, 10 | 1550 | 6.5 kW | 200 fs | 840-3000 | [16] 2013 |
| 17 | SCF | $78TeO_2-5ZnO-12Na_2O-5Bi_2O_3$ | 1560 | 19 kW | 290 fs | 1300-2400 | [17] 2014 |
| 18 | SCF | $76.5TeO_2 -6ZnO-11.5Li_2O-6Bi_2O_3$ | 1958, 2000 | 9.2 kW, 22 kW | 200 fs | 900-3900, 1000-2800 | [18] 2015 |
| 19 | SCF | $70TeO_2-20ZnO-9BaO-BaCl_2$ | 1500, 2400 | ~9 kW | 90 fs | 800- 2800, 1800-3400 | [19] 2015 |
| 20 | T+SCF | $75TeO_2-15ZnO-5ZnF_2-5Na_2O$ | 1560, 2060 | 3 kW | 75 fs, 2.5 ps | 1050-2150, 2100-2700 | [20] 2016 |
| 21 | SI | $80TeO_2-5ZnO-10Na_2O-5ZnF_2$* $60TeO_2-20Na_2O-15GeO_2-5ZnO$** | 2400 | 6 kW | 400 fs | 1300-5300 | [21] 2017 |
| 22 | SI | $80 TeO_2-5ZnO-5ZnF_2-10Na_2O$* $60TeO_2-20Na_2O-15GeO_2-5ZnO$** | 2069 | 600 W | 2.5 ps | 1500-3500 | [22] 2017 |
| 23 | SI | $80TeO_2-5ZnO-10Na_2O-5ZnF_2$ (core) and $60TeO_2-5ZnO\ 20Na_2O-15GeO_2$ (cladding) | 2300 | 3 kW | 200 fs | 1500-4500 | [23] |
| 24 | SI | $80TeO_2-5ZnO-10Na_2O-5ZnF_2$ (core) and $60TeO_2-5ZnO\ 20Na_2O-15GeO_2$ (cladding) | 2000 | 5.5 kW | 200 fs | 1340–2840 | [24] |

In particular, suspended core fibers are attractive for their strong nonlinear response, stemming from high mode confinement. However, an efficient excitation of the fundamental mode in such structures is more challenging, than in fibers with smaller refractive index contrast between the core and cladding. It is also well known that a regular lattice photonic crystal fiber (PCF) can be endlessly single mode [26].

For tellurite glasses, the combination of advantageous features includes their material dispersion, which makes fibers made of these glasses compatible with robust femtosecond fiber-based lasers operating around 1550 nm ($Er^{3+}$-doped fibers) or with the emerging 1900-2100 nm lasers ($Tm^{3+}$ or $Tm^{3+}$-$Ho^{3+}$ doped fibers). This motivates to investigate tellurium oxide glass fibers with engineered all-normal dispersion (ANDi) characteristics. ANDi fibers, combined with femtosecond laser pumping, enable generation of pulse-preserving supercontinuum light [27-29], which offers unique possibilities for further processing, including compression down to single optical cycles [30] or seeding of optical amplifiers and amplification of high quality pulses up to regimes compatible with attosecond science [31].

Here, we report our results on designing, fabrication and characterization of tellurite glass PCFs, which combine the following advantages: high nonlinearity, potential for MIR transmission, normal dispersion profile for coherent, pulse-preserving supercontinuum generation and reasonable mechanical and chemical stability. Optimisation of the basic geometric parameters in a regular hexagonal air-hole lattice [32], including the size of air holes, the pitch and the number of rings, allowed to design a PCF with normal, flattened chromatic dispersion profile up to a wavelength of around 3 μm. The first part of work includes numerical calculations for a hexagonal lattice geometry to obtain all-normal dispersion with a single-mode performance starting at a wavelength of 0.8 μm. Technological limitations are also considered by running simulations for structures with slightly modified air-hole diameter or pitch parameters for different rings. Then, we numerically validate the selected fiber design with nonlinear propagation simulations, which allows to expect octave spanning supercontinuum spectra, delivered by pulses with preserved temporal profiles, and by pumping with commercially available, moderate peak-power, erbium fiber-based femtosecond lasers. This is followed by development of two test series of fibers, which are characterized in terms of chromatic dispersion, measured over 1200-2400 nm wavelengths and supercontinuum generation experiments. Turn-key, fixed wavelength erbium fiber-based femtosecond laser is used for pumping. The supercontinuum spectrum obtained from one of the developed fibers covers spectral range of 1100 nm to 2600 nm. The high nonlinearity of the demonstrated fibers enabled to obtained ANDi supercontinuum spectrum under pumping with a roughly 40 kW peak power, femtosecond fiber-based laser, which in terms of spectral coverage is comparable to, or exceeds the performance possible with previous ANDi fiber designs only under pumping with large optical parametric amplifier systems with disposed peak power in the MW range [27,29]. This is also one of the first tellurite oxide glass, hexagonal air-hole lattice PCFs with an engineered, all-normal dispersion profile, following a similar structure albeit with anomalous dispersion demonstrated previously [14].

## 2. PCF layout design and dispersion engineering

Numerical calculations were performed using parameters of tellurite glass labelled TWPN/I/6 synthesized at Institute of Electronic Materials Technology [14,33]. Material dispersion of tellurite glass is described by the second-order Sellmeier equation:

$$n^2(\lambda) = 1 + \frac{B_1\lambda^2}{\lambda^2-C_1} + \frac{B_2\lambda^2}{\lambda^2-C_2} + \frac{B_3\lambda^2}{\lambda^2-C_3}, \qquad (1)$$

where $B_1$=2.49628, $B_2$=0.79831, $B_3$=1.8848, $C_1$=0.01699 µm², $C_2$=0.07728 µm², $C_3$=162.9949 µm² are the Sellmeier coefficients for the TWPN/I/6 glass. The glass has a nonlinear refractive index value of $n_2$=5.11×10$^{-19}$ m²/W, measured at 1064 nm [34].

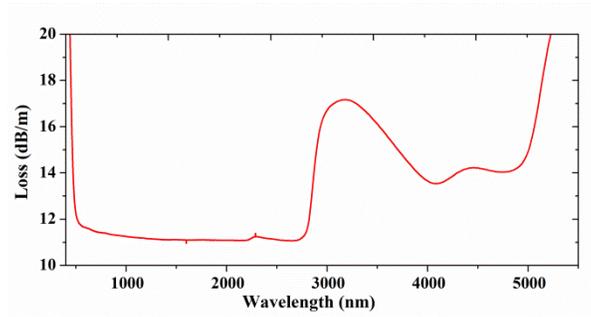

Fig. 1. Attenuation characteristic of the TWPN/I/6 tellurite glass.

Attenuation characteristic of the glass is presented in Fig. 1. The measurement was performed for two samples – 2 mm and 4 mm thick – and the result shown here is corrected for Fresnel reflections. Tellurite glass TWPN/I/6 has reasonably high transmission in the range from 500 nm to 2800 nm. Therefore we limit the spectral range of our optimisation work to 800-3000 nm wavelengths. Material loss is not taken into account during structure optimisation process. We note, that our glass was not purified with respect to OH content, although in general this is technologically feasible and has been successfully demonstrated for tellurite glasses by other groups [16,21].

In the modelling, we consider a PCF geometry with regular hexagonal lattice of air holes. This allows to adjust dispersion and modal properties in a broad wavelength range, as has been demonstrated by Saitoh et al. in the case of silica glass PCFs [32]. Modification of air hole diameters in subsequent rings of the photonic cladding opens an additional degree of freedom in fiber design. Geometrical parameters of the first ring of air holes have the foremost impact on the dispersion profile. However, in practice it is difficult to maintain geometrical parameters of the two subsequent rings with different air hole diameters. In order to anticipate and evaluate potential influence of fabrication accuracy on physical fiber properties i.e. the dispersion profile, we consider the subsequent rings, as well. Therefore our proposed pattern has two different relative air hole sizes ($D_i/\Lambda$ and $D_e/\Lambda$) for internal and external rings respectively as shown in Fig. 2.

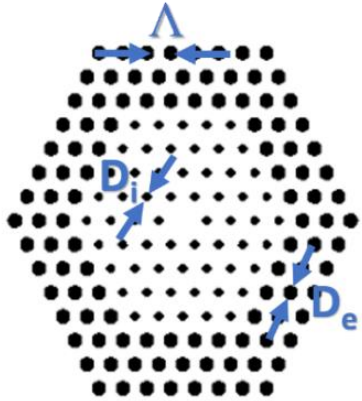

Fig. 2. PCF lattice layout investigated in this work.

The modelling was performed in three stages accordingly to the most important PCF parameters: the number of rings, the diameter of air holes (in internal and external rings) and the lattice pitch $\Lambda$. Firstly, we calculated confinement losses according to the number of rings (Fig. 3) in a PCF with constant air-hole diameters $D_i=D_e$ and for various lattice pitch values.

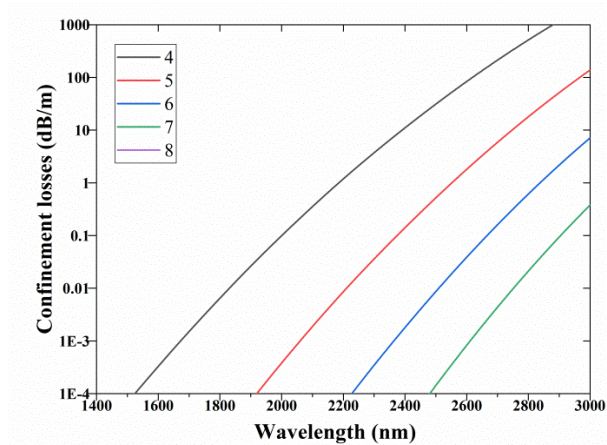

Fig. 3. Confinement losses of fundamental mode for different numbers of rings in a hexagonal double air-hole lattice with parameters: $D_i=D_e=0.8$ μm and $\Lambda=1.2$ μm.

The number of rings in a cladding has influence mainly on confinement losses of the guided modes, because adding more lower refractive index elements prevents spreading of the electric field away from the core. If the number of rings is increasing, then the long-wavelength edge of the fundamental mode confinement loss characteristic shifts towards the higher wavelengths. The purpose of our investigation is to assure efficient supercontinuum (SC) generation up to 2800 nm, thus a 7 ring model is sufficient. For a wavelength of 2800 nm we obtained waveguide losses below 0.01 dB/m (material losses are not considered in this simulation).

The next step of simulations is devoted to the determination of air hole diameters in internal and external rings. We assumed in this simulation a single internal ring of air holes with the same relative air hole size in the range of 0.16-0.5 and 6 external rings with another, similar relative air hole size in the range of 0.58- 0.92. The influence of air hole diameters on dispersion and confinement losses is shown in Fig. 4. Since modification of internal ring diameters is mainly responsible for dispersion profile and of the external rings for attenuation of the whole structure, first we optimised the dispersion characteristics with parameters of air holes in the internal ring. This was followed by optimisation of the waveguide loss of the fiber by modifying parameters of air holes in the external rings.

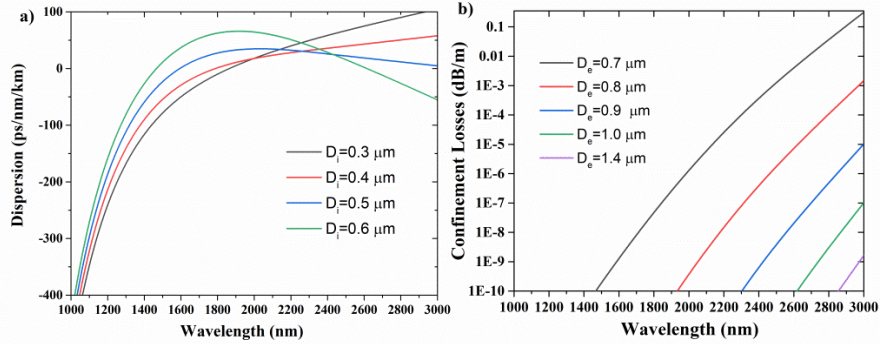

Fig. 4. a) Dispersion profiles for different diameters of a single internal ring $D_i$ with fixed diameter of the 6 external rings $D_e$=0.8 µm, and b) confinement losses for different diameters of the 6 external rings $D_e$ with the internal ring diameter of $D_i$=0.5 µm and for constant $\Lambda$=1.2 µm.

Performed simulations showed that all-normal dispersion is not feasible in PCFs with single internal ring with considered air hole diameters in the range between 0.3 and 0.6 µm, Fig. 4a. Introduction of additional inner rings is required to modify the dispersion profile. Use of multiple internal rings with similar air hole diameter is also beneficial from a technological point of view. Drawing of soft glass PCFs with various air-hole dimeters in close proximity is difficult to maintain to reproduce the original design.

The most flattened dispersion profile in the range of 1600-3000 nm wavelengths is obtained for air holes with diameter of $D_i$=0.5 µm in the internal ring. This structure was selected for further optimisation. As expected, any change of parameters in external rings does not influence significantly the dispersion characteristics, but modifies significantly the confinement loss of the fundamental mode. This loss decreases with increasing air holes diameter. However it is technologically challenging to develop PCF structures with very large relative air hole size in case of tellurite glass [14]. Therefore, use of air holes with diameter $D_e$=0.8 µm is a compromise between confinement losses of 0.001 dB/m for the fundamental mode and the technological limits. Furthermore, it allows to reduce difference in diameters between air holes in internal and external rings ($D_i/D_e$=0.75), which is advantageous from a practical point of view of drawing of this fiber.

In the next step we analyse dispersion characteristics and confinement losses for the 7 ring PCF structure with various number of internal and external rings. The simulations are performed for air hole dimeters of $D_i$=0.5 µm and $D_e$=0.8 µm, in internal and external rings, respectively. We introduced a new parameter of the fiber structures R defined as a ratio of the number of internal rings to the number of external rings $R=n_{Di}/n_{De}$, where $n_{Di}$ and $n_{De}$ denote the number of internal and external rings, respectively. In general, we observe that internal rings are mostly responsible for tailoring of the dispersion and external rings for the reduction of confinement losses, as shown in Fig. 5. We conclude that introduction of a second internal ring (R=2/5) significantly modifies dispersion characteristics of the fiber. Dispersion is reduced below zero, while its flattened profile is preserved. Further increase of the number of internal rings only slightly modifies dispersion characteristics. Simultaneously, the confinement loss for the fundamental mode increases in particular for longer wavelengths. At least 4 external rings are required to maintain confinement losses below 0.1 dB/m for wavelengths below 2400 nm (R=3/4). In conclusion we select, as the most beneficial, the structure with ratio R=4/3, because it offers low confinement losses and flat all-normal dispersion characteristics in particular in the range of 1600-2800 nm wavelengths. The fiber structure with ratio parameter R=2/5 has even lower attenuation, but its dispersion characteristics is more sensitive to any technological inaccuracies. Any discrepancies with respect to the original design in the air hole diameter expected for various air-holes dimeters in close proximity for the 3$^{rd}$ and 4$^{th}$ rings will not influence significantly neither the dispersion characteristics nor the confinement losses.

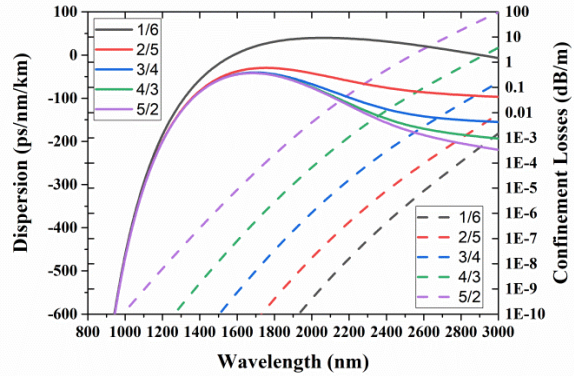

Fig. 5. Dispersion profiles and confinement losses of fundamental mode of designed PCF according to R parameter, assuming holes diameters $D_i$=0.5, $D_e$=0.8 and lattice constant $\Lambda$=1.2 µm.

Finally, we optimise the lattice pitch of the considered PCF with 7 rings in photonic cladding. The relative air hole size of the internal $f_i$=0.42 and external $f_e$=0.67 rings of air holes is maintained in the photonic cladding, as well as the ratio parameter R=4/3 between number of internal and external rings. An overview of how the lattice constant $\Lambda$ influences dispersion and effective mode area is shown in Fig. 6.

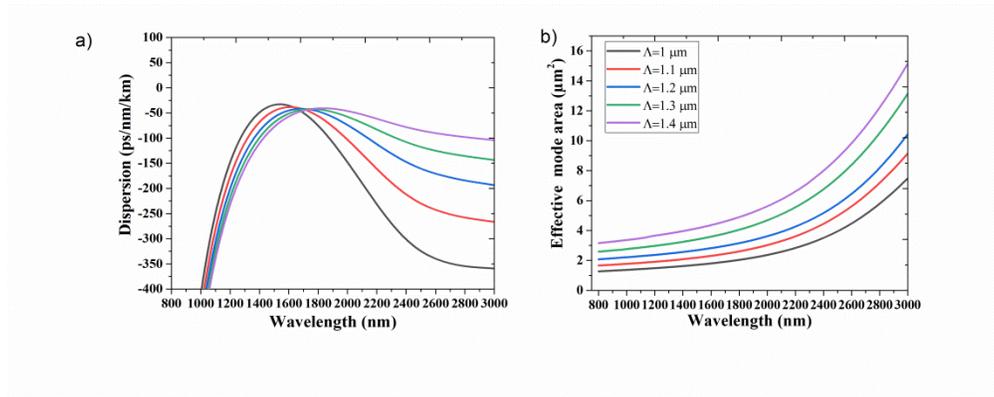

Fig. 6. Influence of lattice constant changes on a) dispersion profile and b) effective mode area of fundamental mode in designed PCF. Rings ratio parameter is R=4/3 and relative air hole sizes of internal and external rings are $f_i$=0.42, $f_e$=0.67, respectively.

Changes of the lattice pitch $\Lambda$ within the considered range between 1.0 and 1.4 µm do not influence confinement losses significantly. The influence on dispersion is mainly noticeable in the longwave range, where the dispersion significantly drops from D=-50 ps/nm/km for $\Lambda$=1.4 µm, down to D=-350 ps/nm/km for $\Lambda$=1.0 µm and for wavelength $\lambda$=3000 nm, as shown in Fig. 3a. Therefore large lattice pitch values are required to maintain flattened normal dispersion. We also calculated the effective mode area for the fundamental mode for the considered lattice constants. Effective mode area increases significantly with the increase of the lattice constant for increasing wavelength, which is disadvantageous for supercontinuum generation, due to decreased nonlinearity of fiber at the redshifted wing and edge of spectrum. We obtained convenient performance in wavelength range of 1.6-3 µm of flattened dispersion profile for 1.3 µm pitch. Dispersion values are changing from -50 to -150 ps/nm/km and effective mode area is below 13 µm² up to the wavelength of 2800 nm. PCF structure optimized with regard to normal dispersion profile and weak dependence of effective mode area on wavelength, is characterized with the following set of geometrical parameters: lattice constant $\Lambda$=1.3 µm, 4 internal rings of the photonic cladding with air holes with the diameter of $D_i$=0.5 µm and 3 external rings with air holes with the diameter of $D_e$=0.8 µm.

In the final step of linear simulations, we verified modal properties of the optimised fiber structure. The confinement losses for the selected fiber structure for the 3 highest order modes are presented in Fig. 7. According to simulation results, the designed fiber has a cut-off wavelength at 2000 nm, when we assume attenuation of 1 dB/m as the guidance limit. However, we can notice that the fundamental and two higher order modes (HOM) are well spatially separated. HOMs are localized in internal rings of the photonic cladding and they can be further attenuated with small bending. Therefore it is possible to excite selectively the fundamental mode which is well confined in the fiber core. The fiber can be considered effectively single mode at the intended pump wavelength of 1560 nm. The nonlinear coefficient of the optimised fiber at this wavelength is $\gamma$ = 582.6 1/W/km.

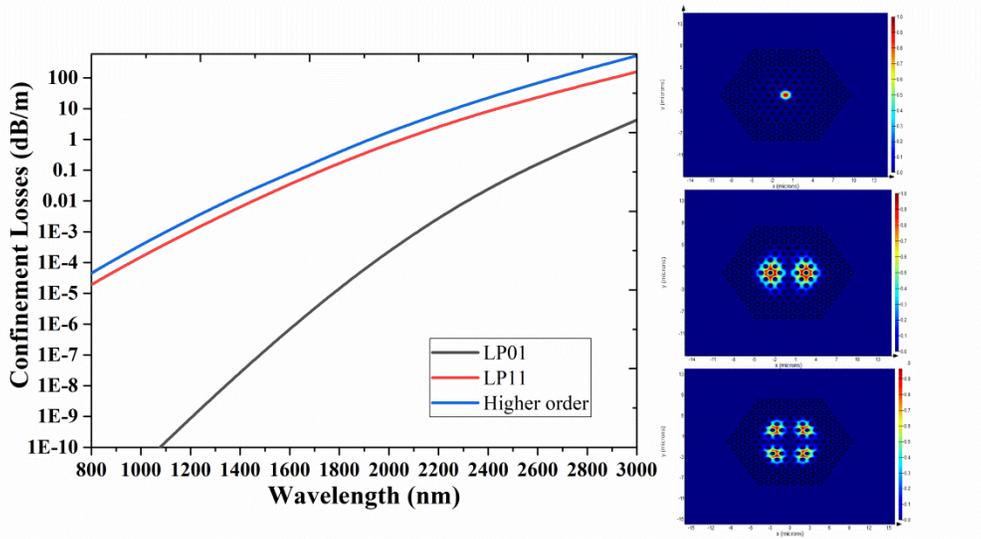

Fig. 7. Confinement losses and electric field distributions for the fundamental mode $LP_{01}$ and two higher order modes in optimized tellurite regular lattice PCF with the lattice constant $\Lambda=1.2$ μm, 4 internal rings of the photonic cladding with air holes with the diameter of $D_i=0.5$ μm and 3 external rings with air holes with the diameter of $D_e=0.8$ μm.

## 3. Nonlinear propagation simulations

The optimized fiber structure has been evaluated for pulse preserving supercontinuum generation performance with nonlinear propagation simulations using the generalized nonlinear Schrödinger equation (GNLSE). We used the model proposed by Travers et al. [35], which we extended to include frequency-dependent loss and effective mode area, as well as one photon per mode noise. Raman parametrization was based on Raman scattering spectrum measurement performed for the used tellurite glass, with the first-order Raman peak located at $\Omega_R=20.4$ THz and its full width at half maximum of 5.26 THz. This corresponds to the Raman parameters used in the modelling of $\tau_1=0.0055$ ps and $\tau_2=0.032$ ps. The Raman contribution to Kerr nonlinearity parameter has been evaluated at $f_R=0.2$, based on fitting of nonlinear soliton self-frequency shift observed in a tellurite PCF with anomalous dispersion and reported previously [14].

Simulations were performed with the assumption of a commercially available fiber-based femtosecond laser as the pump source. Specifically, the pump pulse parameters used in the modelling were taken after Menlo C-Fiber, available from Menlo Systems [36]. This laser delivers Gaussian shaped pulses

$$E(T) = \sqrt{P} * \exp\left(\frac{-T^2}{2*t_0^2}\right), \qquad (2)$$

where P denotes peak power, T – time width, and $t_0$ – duration of input pulse. The remaining pulse parameters were: pulse duration (autocorrelation) of 90 fs, pulse energy up to 5 nJ, the spectrum has been scaled in spectral power density taking into account the laser repetition rate of 100 MHz.

The fiber length assumed in the modelling was 4 cm. With this, the material loss of the fiber can be neglected, while advantage is taken of the fact, that the spectrum develops rapidly upon injecting a femtosecond pulse into the fiber. The obtained spectral broadening and pulse evolution are show in Fig. 8 with color map plots of a spectrogram trace and spectral evolution of the supercontinuum along the fiber. Temporally uniform pulses typical for femtosecond supercontinuum generation over wavelengths with normal dispersion of the nonlinear medium are obtained, and the spectrum spans a full octave already for pump pulse energy of 0.419 nJ. The nonlinear mechanics of supercontinuum development in this case include, as expected, self-phase modulation (SPM) around the central pump wavelength, followed by optical wave breaking (OWB) at the leading and trailing edges of the pulse, and the development of spectrum is completed by onset of the parametric sidebands. These sidebands are the result of mixing, which in the time domain takes place between the SPM spectral components and the pulse components at either the leading or the trailing edge [37]. SPM occurs rapidly over the first millimetres of propagation, next the OWB sidebands can be observed, first at the blue-shifted (trailing) edge of the pulse, followed by the sideband at the red-shifted (leading) edge of the pulse. The red-shifted sideband development is limited by decreasing nonlinearity of the fiber, related to increasing mode area (Fig. 6b). Broadening of spectra toward the shorter wavelengths is limited to 1.0 μm due to steep dispersion characteristics which shortens the time overlap of the mixing components for increasing detuning from the central part of the spectrum (and the pulse), as shown in Fig. 6a. Despite this, an over octave spaning supercontinuum is obtained, covering 1000-2200 nm. Notably, most of the spectral broadening occurs in this simulation over the first 1 cm of propagation.

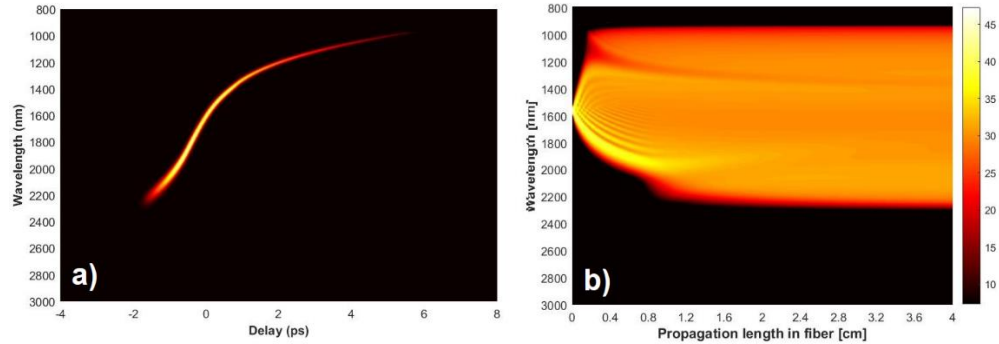

Fig. 8. Evolution of SC spectrum in case of a) pulse delay and b) propagation length along the optimized fiber after 4 cm propagation for 1560 nm laser pump with 1 nJ pulse energy.

Supercontinuum generation performance was then verified numerically assuming only 1 cm long fiber, but for a range of pulse energies – resulting ensemble of spectra is shown in Fig. 9. Full octave spanning spectrum was obtained for in-coupled pulse energy of just below 1 nJ. Assuming a conservative 33% coupling efficiency to a fiber in real experimental conditions, this result would require about 3 nJ of incident pump pulse energy, which is safely within the operational parameters of the commercial laser system referenced earlier [36].

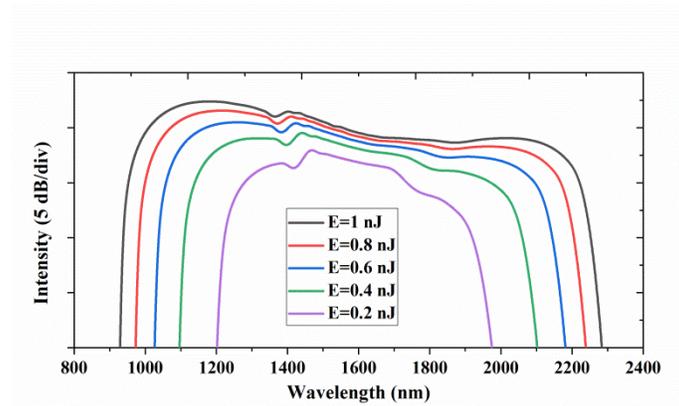

Fig. 9. Numerically calculated SC output spectra for a range of pump pulse energies from a 90 fs laser with central wavelength of 1560 nm.

## 4. Discussion of experimental results with physical fibers

Two test series of tellurite glass photonic crystal fibers have been fabricated, which included fiber samples labelled NL47A2, NL47A3, NL47A4 and NL47B1, NL47B2, NL47B3. A typical image of the fiber structure, here shown for the fiber labelled NL47B1, obtained with scanning electron microscopy (SEM) is presented in Fig 10. Geometrical parameters of the developed fibers, taken from the SEM images, are summarized in Table 2. In some of the fibers, individual defects occurred, eg. in form of collapsed air holes. However, this was observed in the outer ring of air-holes and therefore the influence of these defects on the fundamental mode can be considered negligible. Fiber structure found as optimal in numerical simulations was characterized with relative air-hole size d/Λ of 0.38 in the inner rings of air holes and 0.61 in the outer ring of air holes. The test fibers, developed in reference to the designed structure, represent a range of geometric parameters around this optimal set with reasonable accuracy, considering technological challenge of physically reproducing a soft glass air hole lattice with two different air hole diameters.

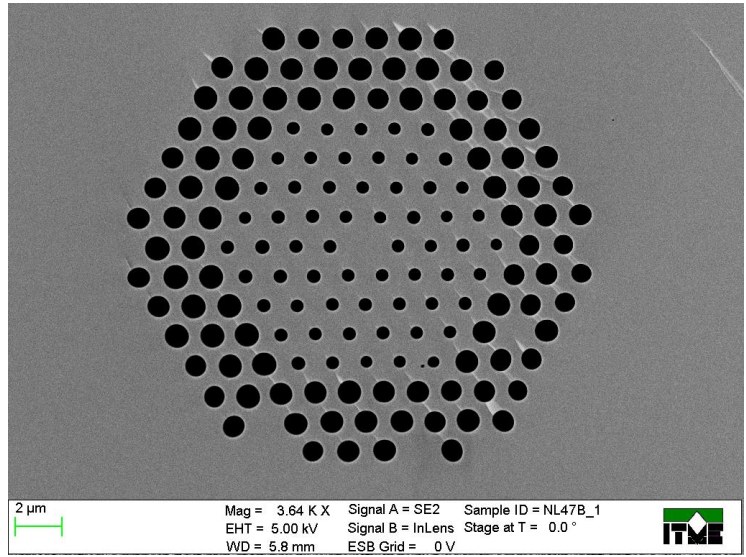

Fig. 10. Typical SEM image of tellurite glass photonic crystal fiber microstructure developed in scope of this work (shown here: fiber NL47B1).

**Table 2. Overview of lattice parameters of the developed tellurite glass photonic crystal fibers.**

| Fiber sample | Fiber diameter (μm) | Photonic cladding diameter (μm) | Core diam. (μm) | Air hole size d (μm) (inner / outer rings) | Relative air hole size d/Λ (inner / outer rings) |
|---|---|---|---|---|---|
| NL47 A2 | 132.7 | 19.96 | 2.16 | 0.66 / 1.06 | 0.46 / 0.73 |
| NL47 A3 | 134.6 | 19.92 | 2.21 | 0.60 / 1.03 | 0.43 / 0.73 |
| NL47 A4 | 154.4 | 22.13 | 2.60 | 0.63 / 1.04 | 0.39 / 0.65 |
| NL47 B1 | 140.3 | 19.87 | 2.32 | 0.57 / 1.00 | 0.39 / 0.68 |
| NL47 B2 | 136.7 | 19.49 | 2.22 | 0.52 / 0.97 | 0.37 / 0.69 |
| NL47 B3 | 135.3 | 18.7 | 2.29 | 0.49 / 0.82 | 0.35 / 0.60 |

Fig. 11-18 contain experimental results on characterization of the developed tellurite glass fibers. The chromatic dispersion measurements were performed using a balanced Mach-Zehnder interferometer setup equipped with a compact supercontinuum source (Leukos) and an optical spectrum analyzer (OSA, Yokogawa, 1200-2400 nm). The same setup was used for dispersion measurements reported eg. in [38]. All supercontinuum spectra were obtained under pumping from a compact erbium fiber-based femtosecond laser from Menlo Systems (model C-fiber HP), which delivers around 90 fs long, nearly transform-limited, Gaussian-shaped pulses, centered at 1560 nm with a repetition rate of 100 MHz. This corresponds to around 40 kW of peak power. Supercontinuum spectra were recorded with a 1200-2400 nm OSA (Yokogawa) in cases where the spectrum was contained within the sensitivity range of this device. In the case of spectra exceeding this range, Thorlabs Fourier Transform analyzer was used with a sensitivity range of 1000-5600 nm (model OSA205) and compared with the diffraction-based OSA measurement within the common wavelength rage. This way, we assured that no Fourier transform errors were recorded in the presented spectra. When discussing the spectral width of supercontinuum spectra recorded in each of the investigated fibers, we assumed a -20 dB reference level, which was feasible with either of the optical

spectrum analyzers used in this work. The femtosecond pump was coupled into the fibers through an aspheric lens (Thorlabs) with a focal length of 4 mm and an anti-reflection coating compatible with the pump laser spectrum. Collimation of light from the PCFs was realized using a similar, but uncoated lens. Fiber lengths used in the experiments varied between roughly a dozen cm to about 30 cm, which was motivated by handling convenience.

The first test series of PCFs, NL47Ax, with x = 2, 3, 4, were fabricated with decreasing relative air hole size in the inner rings, and consequently increasing core size, see Tab. 2. The measured dispersion profiles of the three fibers are shown in Fig. 11 with a tendency of decreasing local maximum with decreasing relative air hole size in the inner rings. Notably, fiber NL47A2 with the largest d/Λ has a zero dispersion wavelength at almost 1600 nm, while the two remaining fibers in the series have all-normal dispersion profiles. Supercontinuum spectra obtained in this series of fibers are shown in Fig. 12 through Fig. 14. Expectedly, the pump power-dependent evolution of spectrum in fiber NL47A1 shows clear solitonic dynamics with increasing incident pump power, as shown in Fig. 12. Measurements of chromatic dispersion and pump power-dependent supercontinuum evolution in the NL47A3 and A4 fibers, each having further decreased value of d/Λ in the inner rings of air holes, confirmed feasibility of obtaining all-normal, concave profiles of dispersion – Fig. 11, as well as corresponding SPM and OWB dominated dynamics of supercontinuum spectral development with increasing pump power – Fig. 13, Fig. 14. Supercontinuum spectrum in Fig. 13, fiber NL47A3, extends outside of the sensitivity range of our OSA. However the heightened intensity level close to 2400 nm is assigned to a second order diffraction artifact, coming from the signal below 1200 nm, which is detected, albeit not displayed by the OSA and is difficult to filter out. Spectrum recorded for the NL47A4 fiber is slightly narrower than that from NL47A3 fiber and is contained within the 1200-2400 nm sensitivity range of the OSA. Out of the two fibers, NL47A4 is closer with its d/Λ to the designed structure. Its nonlinear performance in terms of achievable supercontinuum spectral width, compared to the NL47A3, could be related to lower quality of cleave and reduced pump coupling efficiency, or to somewhat larger core diameter and thus lower nonlinearity. Both chromatic dispersion profiles remain at normal values, but fiber NL47A3 has lower normal dispersion thus providing enhanced conditions for OWB, than the NL47A4 fiber. With this, a second series of test fibers has been fabricated, with the range of technological parameters of the drawing process (preform feed and pull rate, gas pressure applied to the preform) beginning where the NL47A3 and A4 fibers had been obtained.

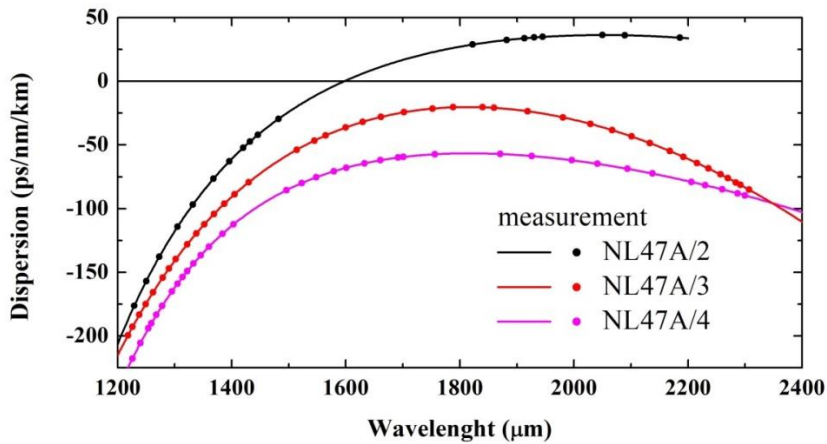

Fig. 11. Chromatic dispersion characteristics measured for NL47A series fibers, dots represent measurement points, solid lines are guides for the eye.

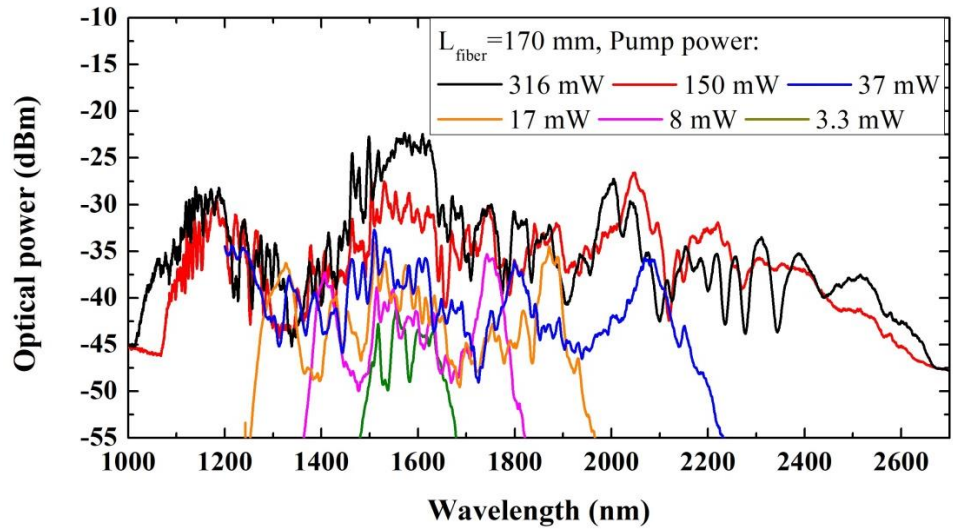

Fig. 12. Supercontinuum spectra measured for the NL47A2 series fibers for different incident average pump powers.

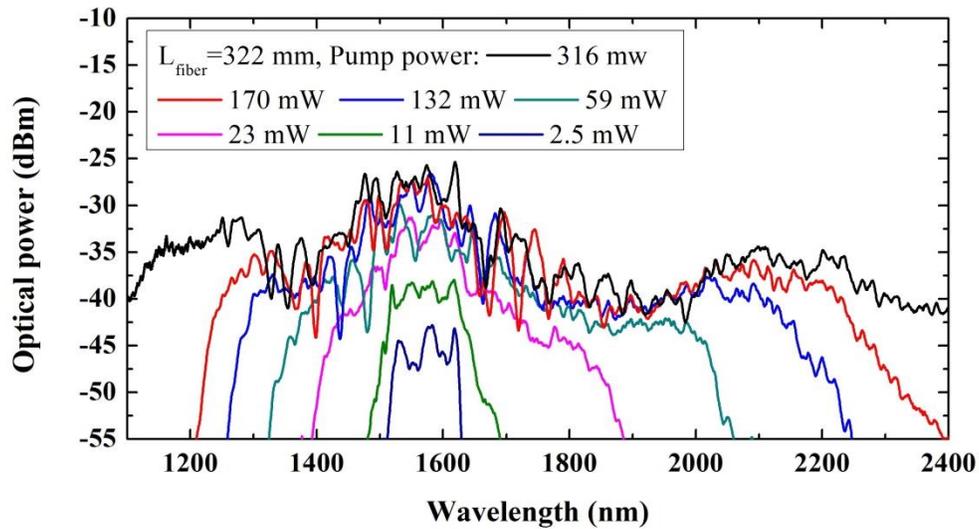

Fig. 13. Supercontinuum spectra measured for the NL47A3 series fibers for different incident average pump powers.

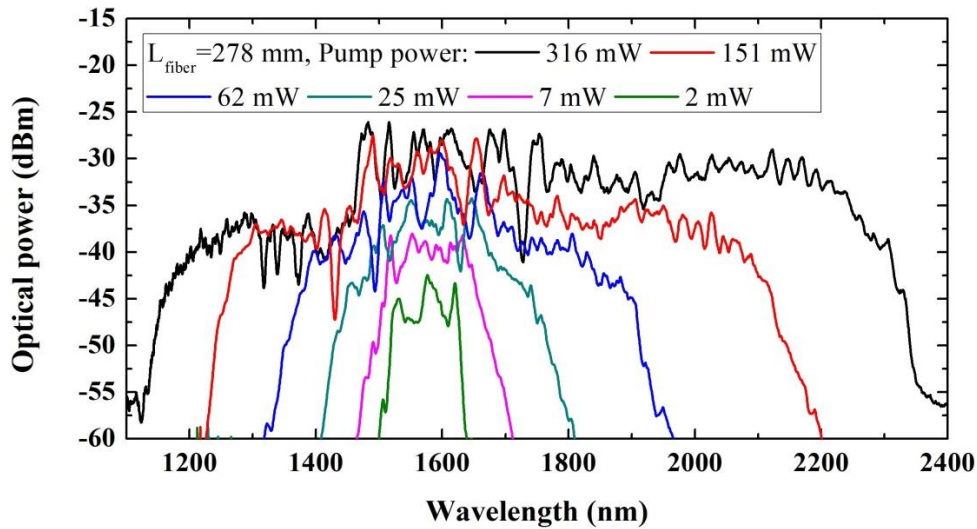

Fig. 14. Supercontinuum spectra measured for the NL47A4 series fibers for different incident average pump powers.

The second test series of fibers, labelled NL47B1 through B3 had relative air hole size of the inner ring between 0.39 and 0.35, thus close to the NL47A3 and A4. Mean core size of the B1-B3 fibers (roughly between 2.3 μm and 2.2 μm of diameter) was just below A4 fiber (2.6 μm). The corresponding, measured dispersion profiles of the B series fibers are shown in Fig. 15. Dispersion of the NL47B2 fiber crosses the zero level at about 1800 nm and the dispersion characteristics of the other two fibers remain at normal values across the wavelength range covered by measurement. In particular, NL47B1 has a normal, concave dispersion profile with a local maximum of around 10 ps/nm/km at about 1800 nm. The largely different dispersion profile of relatively similar fibers B2 and B3 is explained by the fact, that B3 has larger core and smaller air holes in both inner and outer rings, compared to B2. Thus the guided mode in fiber B3 experiences more material dispersion, compared to fiber B2, and material dispersion of the TWPN/I/6 glass is normal at the wavelengths of interest. Data in Tab. A indicates, that the NL47B1 fiber is a geometric compromise between B2 and B3 fibers, and features the largest core diameter of the three samples, the largest air hole dimeters in the cladding and relative air hole size is simultaneously matching well the numerical design proposed for these fibers. Dispersion measurement results obtained for this test series of fibers have been supported with numerical simulations, based on SEM images of real fiber structures. The calculated dispersion profiles, shown in Fig. 15 with dotted traces for each of the fibers, are in good agreement for fibers NL47B1 and B3. There is a significant mismatch between measurement and simulation for the NL47B2 fiber. This finds explanation in the fact, that dispersion measurement based on balanced Mach-Zehnder interferometer makes it difficult to follow the zero-order interferometric fringe for very small dispersion values and the redshifted part of the measured dispersion profile is in fact, characterized by limited accuracy, although both experimentally and numerically obtained data suggests existence of at least one ZDW for this fiber. The supercontinuum spectra obtained in all three fibers are shown in Fig. 16 through Fig. 18. For cases, where the spectrum exceeded the 1200-2400 nm sensitivity range of our diffraction-based OSA, the Thorlabs OSA205 Fourier transform OSA was used, hence also a noticeable difference in the noise floor for these measurements. High nonlinearity relevant to the tellurite glass and very favorable dispersion profile facilitated a largest spectral width of supercontinuum spectrum obtained in the NL47B1 fiber, spanning a wavelength range from 1100 nm up to almost 2600 nm, among all fibers investigated in this work. This result exceeds the expectations for this fiber, based on the numerical simulations discussed in the previous section. We note, that in our numerical simulations a coupling efficiency of 33% was assumed, which was both conservative and just enough to generate a full octave supercontinuum, considering the pump parameters assumed in the modelling. Exceeding of the numerical result with the physical experiment in our work is explained by the fact, that in experimental conditions. coupling efficiency in the order of 50-60% is readily achievable with fiber-based mode-locked lasers, provided that the fiber's numerical aperture (NA) is matched by the NA of the in-coupling aspheric lens. Pump power-dependent evolution of spectrum recorded for this fiber confirms the SPM/OWB features, with the characteristic asymmetry over the wavelength scale attributable to steep normal dispersion slope at blue-shifted wavelengths (from the pump wavelength). Combination of high nonlinearity and low dispersion of nonlinearity, with low value of normal dispersion across the redshifted wavelength range support efficient spectral broadening towards 2600 nm. Small absolute value of dispersion in the NL47B2 fiber masks solitonic features in the pump power-dependent dynamics of spectral broadening. The B2 fiber is similar in terms of absolute dispersion and geometric dimensions to the B1 fiber. Despite careful fiber cleaving, the B2 fiber did not allow to duplicate the record spectral width of the B1 fiber. In light of this, as well as taking into account the clear indication of a ZDW presence by both experimental and theoretical data for the B2 fiber, further supported by lack of evidence for clear soliton dynamics, allows to conclude that the dispersion profile of the NL47B2 fiber has indeed two closely located ZDWs. Large value of normal dispersion in case of NL47B3 fiber results in limited spectral broadening, similarly to the NL47A4 fiber.

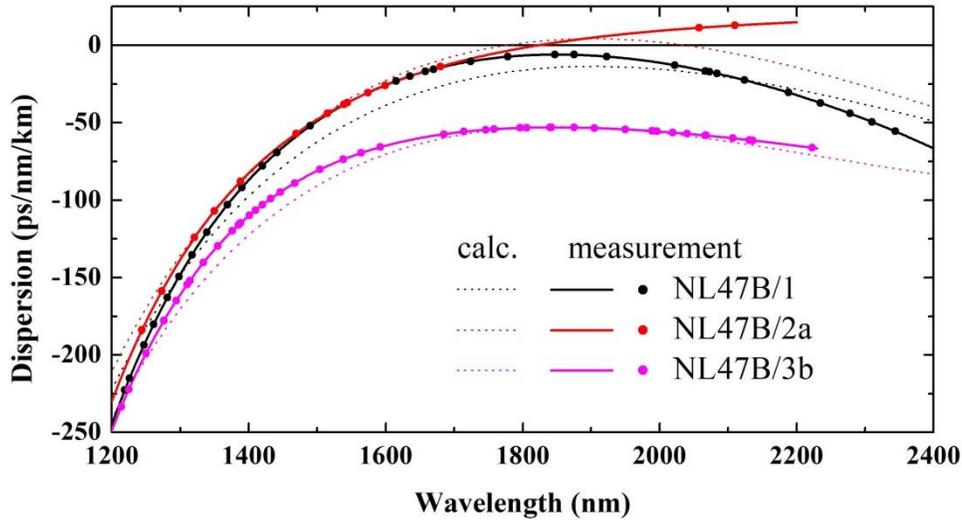

Fig. 15. Chromatic dispersion characteristics measured and calculated for NL47B series fibers. Dots represent measured data and solid lines are guides for the eye, dotted traces are calculated dispersion profiles.

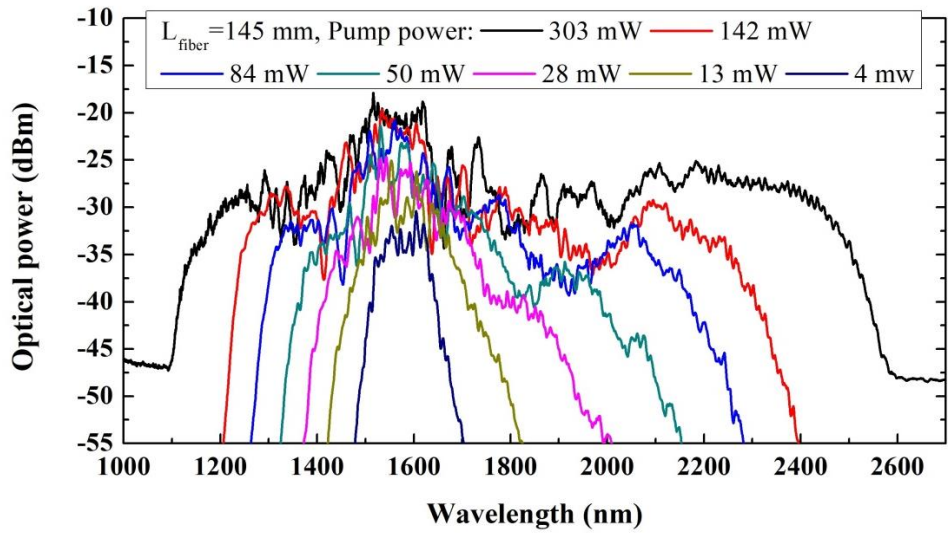

Fig. 16. Supercontinuum spectra measured for the NL47B1 fiber for different incident average pump powers.

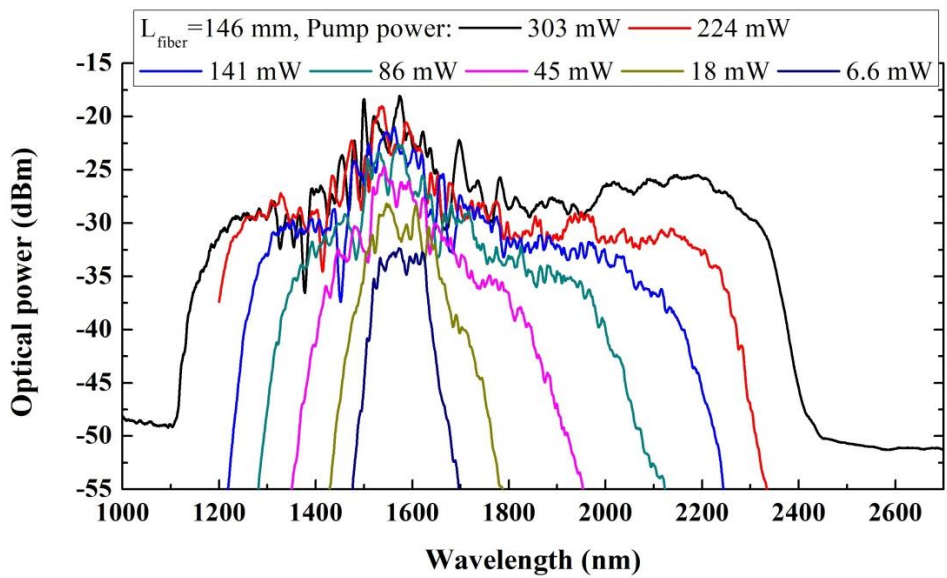

Fig. 17. Supercontinuum spectra measured for the NL47B2 fiber for different incident average pump powers.

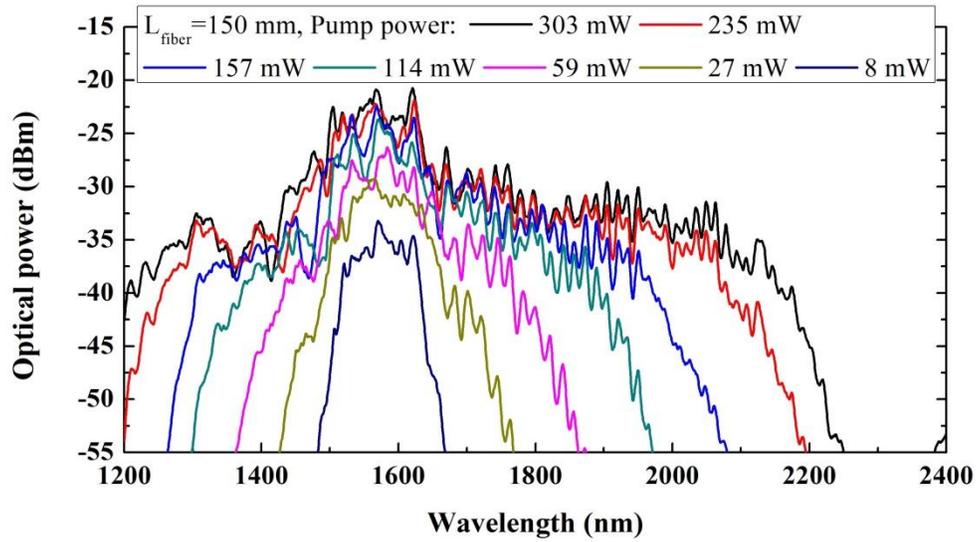

Fig. 18. Supercontinuum spectra measured for the NL47B3 fiber for different incident average pump powers.

It is to be stressed, that the discussed results of supercontinuum generation were obtained using a fixed wavelength, fiber-based femtosecond laser operating at 1560 nm. As said, the laser delivered 90 fs long Gaussian pulses, which is very favorable for pulse preserving, ANDi supercontinuum generation, albeit its peak power is limited at a level of about 40 kW. This is significantly below 100-200 kW range readily available from 50-60 fs pulsed optical parametric oscillator (OPO) systems very often used in ANDi supercontinuum generation experiments, and specifically used in our previous experiments with other soft glass ANDi PCFs [27,39]. Fig. 19 presents an overview of supercontinuum generation performance of the NL47B1 fiber and the two fibers investigated previously under OPO pumping, this time under pumping from the fixed-wavelength, erbium fiber-based laser. Table 3 contains a summary of the compared fibers, including type of their structure and type of used glasses and their optical parameters relevant to the nonlinear response. This comparison provides a dramatic rationale for the use of highly nonlinear, dispersion engineered soft glass photonic crystal fibers for applications in generation of ANDi femtosecond supercontinuum, when the pump laser economy and its small foot print are issues to be considered. This is of particular importance for recently discussed applications of ANDi supercontinuum in seeding of femtosecond fiber amplifiers [31,40]. The demonstrated tellurite glass ANDi PCFs also extend the current state-of-the-art in fibers designed for pulse-preserving, femtosecond supercontinuum generation, under pumping with lasers operating at standard wavelengths. Up to date, this line-up included fibers made of silica and germanium-doped silica fibers [28,29] and highly nonlinear silicate fibers [39], as well as ANDi chalcogenide glass fibers and waveguides, enabling at least an octave of normal dispersion supercontinuum under femtosecond pumping around 2070 nm [41] or at around 2700-2800 nm [42-45].

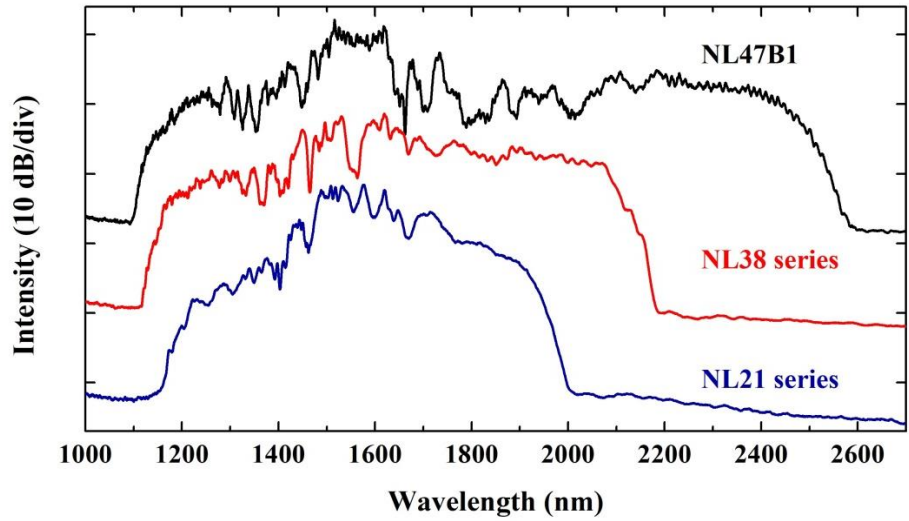

Fig. 19. Supercontinuum spectra obtained under pumping from a robust erbium fiber-based femtosecond laser from three different soft glass photonic crystal fibers: NL21 series all-solid glass PCF made of boron silicate and lead silicate glasses [27], NL38 series all-solid glass PCF made of Schott silicate glasses F2 and SF6 [39], NL47B1 – the most efficient air-hole lattice, tellurite glass ANDi PCF reported in this work.

Table 3. Comparison of nonlinear optical properties of selected heavy-metal oxide soft glass ANDi PCFs.

| | PCF series and type | Nonlinear refractive index of glass in the core | Effective mode area calculated at $\lambda = 1560$ nm | Nonlinear coefficient $\gamma$ at $\lambda = 1560$ nm | Reference |
|---|---|---|---|---|---|
| 1 | NL21 series, all-solid glass PCF, in-house boron-silicate glass and Schott F2 glass (including core) | $2.9 \times 10^{-20}$ m$^2$/W | 6.2 μm$^2$ | 13.6 W$^{-1}$km$^{-1}$ | [27] |
| 2 | NL38 series, all-solid glass PCF, Schott F2 and Schott SF6 glasses (including core) | $21.0 \times 10^{-20}$ m$^2$/W | 4.0 μm$^2$ | 213.7 W$^{-1}$km$^{-1}$ | [39] |
| 3 | NL47 series, air-hole hexagonal lattice PCF, tellurite glass | $51.0 \times 10^{-20}$ m$^2$/W | 3.5 μm$^2$ | 582.6 W$^{-1}$km$^{-1}$ | This work |

## 5. Conclusions

All-normal dispersion supercontinuum generation with femtosecond lasers can find practical application in ultrashort pulse generation and amplification, because femtosecond lasers based on mode-locked fiber oscillators are becoming mature and price-accessible. It is however necessary to develop optical fibers with high nonlinearity, which would compensate for limited peak power of such pump lasers, compared to peak power available from complex and expensive OPO systems. We have investigated a new type of tellurite glass PCFs with engineered ANDi dispersion profiles. Such fibers, combined with femtosecond laser pumping can deliver spectrally broad, coherent light pulses. In selecting of the tellurite glass for fiber development, we were motivated by a combination of convenient features of this material – its high nonlinearity with $n_2 = 51.0 \times 10^{-20}$ m$^2$/W, good power handling at this wavelength range and reasonably good mechanical and chemical properties of the glass. In designing of the fiber lattice, we took the approach with varying air hole diameter in subsequent rings, such that the four inner rings surrounding the core had smaller diameters, than the air hole diameters in the outer three rings of the photonic cladding. In this approach, the internal rings determine dispersion properties of the fiber while external rings define modal properties and confinement losses of the fiber modes. This way we were able to arrive at a dispersion profile flattened over normal values in the range of -10 to -50 ps/nm/km over wavelengths from 1500 nm to 2400 nm in a physically developed PCF. At the intended pump center wavelength of 1560 nm, the fiber supports 3 guided modes, but the fundamental mode is well confined in the fiber core and it is spatially separated from HOMs, which are located in the photonic cladding. As a result, the fundamental mode can be excited selectively and the fiber can be considered effectively single mode. The effective mode area at the intended pump central wavelength of 1560 nm was 3.5 μm$^2$, which corresponded to nonlinear coefficient of $\gamma = 582.6$ W$^{-1}$km$^{-1}$. Among the fibers physically developed in scope of this work, the fiber labelled NL47B1 with relative air hole size d/Λ of 0.39 in the inner rings and 0.68 in the outer rings (matching the design obtained numerically) has been found the best in terms of nonlinear performance. Spectrum of the supercontinuum pulses generated in this fiber covered wavelengths from 1100 nm to 2600 nm, which by far is the broadest spectrum generated in a nonlinear PCF with normal dispersion profile under pumping from a fixed-wavelength femtosecond fiber-based laser. The fiber could find application in coherent seeding of ultrafast fiber amplifiers [31,40], because its high nonlinearity would allow to soften the requirement on high peak power of the 1560 nm femtosecond front-end laser used for pumping of the ANDi supercontinuum seed signal generation. Such systems would additionally benefit from a linearly polarized seed signal and indeed, polarization maintaining ANDi PCFs compatible with 1560 nm femtosecond lasers have been recently reported [46,47]. Therefore, future work on the development of the tellurite glass PCFs reported here, would include modification of lattice parameters to obtain dispersion profiles compatible with pumping from the emerging thulium fiber-based femtosecond lasers at around 2000 nm, as well as on feasibility of introducing birefringence in order to control polarization of the supercontinuum pulses from these fibers.

**Funding.** European Training Network H2020-MSCA-ITN-2016 Grant No 722380, SUPUVIR: Supercontinuum broadband light sources covering UV to IR applications; First TEAM/2016-1/1 operated within the Foundation for Polish Science Team Programme co-financed by the European Regional Development Found under Smart Growth Operational Programme (SG OP), Priority Axis IV.

**Acknowledgment.** The authors acknowledge efforts of Iurii Venglovskyi in modelling tellurite based photonic crystal fibers.

**References**

1. P. Domachuk, M. A. Wolchover, M. Cronin-Golomb, A. Wang, A.K. George, C.M.B. Cordeiro, J.C. Knight, F.G. Omenetto, "Over 400 nm bandwidth of mir-ir supercontinuum generation in sub-centimeter segments of highly nonlinear tellurite PCFs", Opt. Express 16(10), 7161-7168 (2008).
2. X. Feng, W.H. Loh, J. C. Flanagan, A. Camergo, S. Dasgupta, P. Petropoulos, P. Horak, K. E. Frampton, N. M. White, J. H. Price, H.N. Rutt, and D.J. Richardson, "Single-mode tellurite glass holey fiber with extremely large mode area for infrared nonlinear applications," Opt. Express 16(18), 13651–13656 (2008).
3. M. Liao, C. Chaudhari, G. Qin, X. Yan, T. Suzuki, Y. Ohishi, "Tellurite microstructure fibers with small hexagonal core for supercontinuum generation," Opt. Express 17(14), 12174-12182 (2009).
4. M. Liao, X. Yan, G. Qin, C. Chaudhari, T. Suzuki, T. Ohishi, „A highly non-linear tellurite microstructure fiber with multi-ring holes for supercontinuum generation," Opt. Express 17(18), 15481-15490 (2009).
5. X. Yan, G. Qin, M. Liao, T. Suzuki, A. Mori, and Y. Ohishi, "Soliton source generation in a few-mode tellurite microstructure fiber," Appl. Phys. Lett. 95, 171107 (2009).
6. G. Qin, M. Liao, C. Chaudhari, X. Yan, C. Kito, T. Suzuki, and Y. Ohishi, "Second and third harmonics and flattened supercontinuum generation in tellurite microstructured fibers," Opt. Lett. 35(1), 58–60 (2010).
7. I. Savelii, J. C. Jules, G. Gadret, B. Kibler, J. Fatome, M. El-Amraoui, N. Manikandan, X. Zheng, F. Désévédavy, J. M. Dudley, J. Troles, L. Brilland, G. Renversez, and F. Smektala, "Suspended core tellurite glass optical fibers for infrared supercontinuum generation," Opt. Mat. 33(11), 1661-1666 (2011).
8. Z. Duan, M. Liao, X. Yan, C. Kito, T. Suzuki, T. Ohishi, "Tellurite composite microstructured optical fibers with tailored chromatic dispersion for nonlinear applications, Appl. Phys. Exp. 4(7), 072502 (2011).
9. M. Liao, G. Qin, X. Yan, T. Suzuki, Y. Ohishi, "A tellurite nanowire with long suspended struts for low-threshold single-mode supercontinuum generation," J. Light. Technol. 29(2), 194 – 199 (2011).
10. M. Liao, W. Gao, Z. Duan, X. Yan, T. Suzuki, Y. Ohishi, "Directly draw highly nonlinear tellurite microstructured fiber with diameter varying sharply in a short fiber length," Opt. Express 20(2), 1141–1150 (2012).
11. I. Savelii, O. Mouawad, J. Fatome, B. Kibler, F. Désévédavy, G. Gadret, J. C. Jules, P. Y. Bony, H. Kawashima, W. Gao, T. Kohoutek, T. Suzuki, Y. Ohishi, and F. Smektala, "Mid-infrared 2000-nm bandwidth supercontinuum generation in suspended-core microstructured sulphide and tellurite optical fibers," Opt. Express 20(24), 27083-27093 (2012).
12. M. Liao, W. Gao, Z. Duan, X. Yan, T. Suzuki, and Y. Ohishi, "Supercontinuum generation in short tellurite microstructured fibers pumped by a quasi-cw laser," Opt. Lett. 37(11), 2127–2129 (2012).
13. L. Liu, Q. Tian, M. Liao, D. Zhao, G. Qin, Y. Ohishi, and W. Qin, "All-optical control of group velocity dispersion in tellurite photonic crystal fibers," Opt. Lett. 37(24), 5124–5126 (2012).
14. M. Klimczak, G. Stępniewski, H. Bookey, A. Szolno, R. Stępień, D. Pysz, A. Kar, A. Waddie, M. R. Taghizadeh, R. Buczyński, „Broadband infrared supercontinuum generation in hexagonal-lattice tellurite photonic crystal fiber with dispersion optimized for pumping near 1560 nm," Opt. Lett. 38(22), 4679-4682 (2013).
15. G. Wang, T. Jiang, C. Li, H. Yang, A. Wang, Z. Zhang, "Octave-spanning spectrum of femtosecond Yb:fiber ring laser at 528 MHz repetition rate in microstructured tellurite fiber," Opt. Express 21(4), 4703-4708 (2013).
16. I. Savelii, F. Désévédavy, J. C. Jules, G. Gadret, J. Fatome, B. Kibler, H. Kawashima, Y. Ohishi, F. Smektala, "Management of OH absorption in tellurite optical fibers and related supercontinuum generation," Opt. Mat. 35(8), 1595–1599 (2013).
17. Z. X. Jia, L. Liu, C. F. Yao, G. S. Qin, Y. Ohishi, and W. P. Qin, "Supercontinuum generation and lasing in thulium doped tellurite microstructured fibers," Journal of Appl. Phys. 115, 063106 (2014).
18. T. Cheng, L. Zhang, X. Xue, D. Deng, T. Suzuki, and Y. Ohishi, "Broadband cascaded four-wave mixing and supercontinuum generation in a tellurite microstructured optical fiber pumped at 2 μm," Opt. Express 23(4), 4125–4134 (2015).
19. M. Belal, L. Xu, P. Horak, L. Shen, X. Feng, M. Ettabib, D. J. Richardson, P. Petropoulos, and J. H. V. Price, "Mid-infrared supercontinuum generation in suspended core tellurite microstructured optical fibers," Opt. Lett. 40(10), 2237-2240 (2015).
20. C. Strutynski, J. Picot-Clémente, F. Désévédavy, J. C. Jules, G. Gadret, B. Kibler, F. Smektala, "Compact supercontinuum sources based on tellurite suspended core fibers for absorption spectroscopy beyond 2 μm, " Laser Phys. Lett. 13, 075402 (2016).
21. S. Kedenburg, C. Strutynski, B. Kibler, P. Froidevaux, F. Désévédavy, G. Gadret, J. C. Jules, T. Steinle, F. Mörz, A. Steinmann, H. Giessen, and F. Smektala, "High repetition rate mid-infrared supercontinuum generation from 1.3 to 5.3 µm in robust step-index tellurite fibers," J. Opt. Soc. Am. B 34(3), 601-607 (2017).
22. C. Strutynski, P. Froidevaux, F. Désévédavy, J. C. Jules, G. Gadret, A. Bendahmane, K. Tarnowski, B. Kibler, F. Smektala, "Tailoring supercontinuum generation beyond 2 μm in step-index tellurite fibers," Opt. Lett. 42(2), 247-250 (2017).
23. P. Froidevaux, A. Lemière, B. Kibler, F. Désévédavy, P. Mathey, G. Gadret, J.-Ch. Jules, K. Nagasaka, T. Suzuki, Y. Ohishi, and F. Smektala, "Dispersion-Engineered Step-Index Tellurite Fibers for Mid-Infrared Coherent Supercontinuum Generation from 1.5 to 4.5 μm with Sub-Nanojoule Femtosecond Pump Pulses," Appl. Sci. 8(10), 1875 (2018).
24. T. S. Saini, T. H. Tuan, L. Xing, N. P. T. Hoa, T. Suzuki and Y. Ohishi, "Coherent mid-infrared supercontinuum spectrum using a step-index tellurite fiber with all-normal dispersion," Appl. Phys. Express 11, 102501 (2018).

25. F. Désévédavy, G. Gadret, J. C. Jules, B. Kibler, and F. Smektala, "Supercontinuum generation in tellurite optical fibers," in Technological Advances in Tellurite Glasses, V. Riviera, D. Manzani ed. (Springer, 2017), pp. 278-297.
26. T.A. Birks, J. C. Knight, P. J. Russell, "Endlessly single-mode photonic crystal fiber," Opt. Lett. 22(13), 961-963 (1997).
27. M. Klimczak, B. Siwicki, P. Skibinski, D. Pysz, R. Stępien, A. Heidt, C. Radzewicz, R. Buczynski, "Coherent supercontinuum generation up to 2.3 μm in all-solid soft-glass photonic crystal fibers with flat all-normal dispersion," Opt. Express 22(15), 18824-18832 (2014).
28. A. M. Heidt, A. Hartung, and H. Bartelt, "Generation of ultrashort and coherent supercontinuum light pulses in all-normal dispersion fibers," in The Supercontinuum Laser Source, R. Alfano, ed. (Springer, 2016), pp. 247–280.
29. K. Tarnowski, T. Martynkien, P. Mergo, K. Poturaj, G. Sobon, and W. Urbanczyk, ´ "Coherent supercontinuum generation up to 2.2 μm in an all-normal dispersion microstructured silica fiber," Opt. Express 24(26), 30523–30536 (2016).
30. S. Demmler, J. Rothhardt, A. M. Heidt, A. Hartung, E. G. Rohwer, H. Bartelt, J. Limpert, A. Tünnermann, "Generation of high quality, 1.3 cycle pulses by active phase control of an octave spanning supercontinuum," Opt. Express 19(21), 20151 – 20158 (2011).
31. J. Rothhardt, S. Demmler, S. Hädrich, J. Limpert, A. Tünnermann, "Octave-spanning OPCPA system delivering CEP-stable few-cycle pulses 22 W of average power at 1 MHz repetition rate," Opt. Express 20(10), 10870-10878 (2012).
32. K. Saitoh, M. Koshiba, T. Hasegawa, E. Sasaoka, „Chromatic dispersion control in photonic crystal fibers: application to ultra-flattened dispersion," Opt. Express 11(8), 843-852 (2003).
33. R. Stepien, R. Buczynski, D. Pysz, I. Kujawa, A. Filipkowski, M. Mirkowska, and R. Diduszko, „Development of thermally stable tellurite glasses designed for fabrication of microstructured optical fibers," Journal of Non-Cryst. Solids 357(3), 873-883 (2011).
34. J. Cimek, N. Liaros, S. Couris, R. Stępień, M. Klimczak, R. Buczyński, "Experimental investigation of the nonlinear refractive index of various soft glasses dedicated for development of nonlinear photonic crystal fibers," Opt. Mat. Exp. 7(10), 3471-3483 (2017).
35. J.C. Travers, M.H. Frosz, and J.M. Dudley, in "Supercontinuum Generation in Optical Fibers," J.M. Dudley and R. Taylor, eds. (Cambridge University, 2010).
36. http://www.menlosystems.com
37. A.M. Heidt, A. Hartung, G.W. Bosman, P. Krok, E.G. Rohwer, H. Schwoerer, and H. Bartelt, "Coherent octave spanning near-infrared and visible supercontinuum generation in all-normal dispersion photonic crystal fibers," Opt. Express 19(4), 3775-3787 (2011).
38. P Ciąćka, A Rampur, A Heidt, T Feurer and M Klimczak, "Dispersion measurement of ultra-high numerical aperture fibers covering thulium, holmium, and erbium emission wavelengths," J. Opt. Soc. Am. B 35(6), 1301-1307 (2018).
39. M. Klimczak, B. Siwicki, B. Zhou, M. Bache, D. Pysz, O. Bang, and R. Buczyński, "Coherent supercontinuum bandwidth limitations under femtosecond pumping at 2 μm in all-solid soft glass photonic crystal fibers," Opt. Express 24(26), 29406-29416 (2016).
40. J. M. Hodasi, A. Heidt, M. Klimczak, B. Siwicki, and T. Feurer, "Femtosecond seeding of a Tm-Ho fiber amplifier by a broadband coherent supercontinuum pulse from an all-solid all-normal photonic crystal fiber," 2017 European Conference on Lasers and Electro-Optics and European Quantum Electronics Conference (Optical Society of America, 2017), paper CJ_P_7.
41. S. Xing, S. Kharitonov, J. Hu, and Camille-Sophie Brès, "Linearly chirped mid-infrared supercontinuum in all-normal-dispersion chalcogenide photonic crystal fibers," Opt. Express 26(15), 19627-19636 (2018).
42. L. Liu, T. Cheng, K. Nagasaka, H. Tong, G. Qin, T. Suzuki, and Y. Ohishi "Coherent mid-infrared supercontinuum generation in all-solid chalcogenide microstructured fibers with all-normal dispersion," Opt. Lett. 41(2), 392-395 (2016).
43. K. Nagasaka, L. Liu, T. H. Tuan, T. Cheng, M. Matsumoto, H. Tezuka, T. Suzuki and Y. Ohishi, "Supercontinuum generation in chalcogenide double-clad fiber with near zero-flattened normal dispersion profile," J. Opt. 19 095502 (2017).
44. K. Nagasaka, T. H. Tuan, T. Cheng, M. Matsumoto, H. Tezuka, T. Suzuki, and Y. Ohishi, "Supercontinuum generation in the normal dispersion regime using chalcogenide double-clad fiber," Appl. Phys. Express 10, 032103 (2017).
45. T. S. Saini, N. P. T. Hoa, K. Nagasaka, X. Luo, T. H. Tuan, T. Suzuki, and Y. Ohishi, "Coherent midinfrared supercontinuum generation using a rib waveguide pumped with 200 fs laser pulses at 2.8 μm," Appl. Opt. 57(70), 1689-1693 (2018).
46. K. Tarnowski, T. Martynkien, P. Mergo, K. Poturaj, A. Anuszkiewicz, P. Béjot, F. Billard, O. Faucher, B. Kibler, and W. Urbanczyk, "Polarized all-normal dispersion supercontinuum reaching 2.5 μm generated in a birefringent microstructured silica fiber," Opt. Express 25(22), 27452-27463 (2017).
47. D. Dobrakowski, A. Rampur, G. Stepniewski, A. Anuszkiewicz, J. Lisowska, D. Pysz, R. Kasztelanic, and M. Klimczak, "Development of highly nonlinear polarization maintaining fibers with normal dispersion across entire transmission window," J. Opt. (2018) https://doi.org/10.1088/2040-8986/aaf4af